\documentclass[preprint,12pt]{elsarticle}

\usepackage{amssymb}
\usepackage{amsmath}
\usepackage{graphicx} 
\usepackage{xcolor} 
\usepackage{algorithm} 
\usepackage{algpseudocode} 
\usepackage{booktabs} 
\usepackage[pdfencoding=auto, psdextra]{hyperref} 
\usepackage{multirow}
\usepackage{threeparttable} 
\usepackage{subcaption}

\hypersetup{
    colorlinks=true,
    linkcolor=blue,
    filecolor=magenta,      
    urlcolor=blue,
    citecolor=blue,
}

\journal{Information Sciences}

\begin{document}

\begin{frontmatter}


\title{FCOC: A Fractal-Chaotic Co-driven Framework for Financial Volatility Forecasting}

\author[1]{Yilong Zeng}

\author[3,1]{Boyan Tang}

\author[1]{Xuanhao Ren}

\author[1,2]{Sherry Zhefang Zhou\corref{cor1}}
\ead{sherryzhou@uic.edu.cn}

\author[3]{Jianghua Wu}

\author[1,2]{Raymond Lee\corref{cor1}}
\ead{raymondshtlee@uic.edu.cn}
\cortext[cor1]{Corresponding author}

\affiliation[1]{organization={Faculty of Science and Technology, Beijing Normal-Hong Kong Baptist University},
                 city={Zhuhai},
                 postcode={519087}, 
                 country={China}}
\affiliation[2]{organization={Guangdong Provincial Key Laboratory of Interdisciplinary Research and Application for Data Science, Beijing Normal-Hong Kong Baptist University},
                 city={Zhuhai},
                 postcode={519087}, 
                 country={China}}
\affiliation[3]{organization={The Shenzhen Research Institute of Big Data, The Chinese University of Hong Kong, Shenzhen},
                 city={Shenzhen},
                 postcode={518000}, 
                 country={China}}


\begin{abstract}
This paper introduces the Fractal-Chaotic Oscillation Co-driven (FCOC) framework, a novel paradigm for financial volatility forecasting that systematically resolves the dual challenges of feature fidelity and model responsiveness. FCOC synergizes two core innovations: our novel Fractal Feature Corrector (FFC), engineered to extract high-fidelity fractal signals, and a bio-inspired Chaotic Oscillation Component (COC) that replaces static activations with a dynamic processing system. Empirically validated on the S\&P 500 and DJI, the FCOC framework demonstrates profound and generalizable impact. The framework fundamentally transforms the performance of previously underperforming architectures, such as the Transformer, while achieving substantial improvements in key risk-sensitive metrics for state-of-the-art models like Mamba. These results establish a powerful co-driven approach, where models are guided by superior theoretical features and powered by dynamic internal processors, setting a new benchmark for risk-aware forecasting.
\end{abstract}

\begin{highlights}
\item Proposes a novel FCOC framework to resolve dual bottlenecks in volatility forecasting.
\item Introduces a Fractal Feature Corrector (FFC) for high-fidelity market signals.
\item Deploys a Chaotic Oscillation Component (COC) to resolve model complexity mismatch.
\item Establishes a co-driven paradigm synergizing fractal features and chaotic dynamics.
\end{highlights}


\end{frontmatter}


\section{Introduction}
\label{sec:introduction}
The forecasting of financial market volatility, as the second moment of asset returns, is a cornerstone of modern finance, crucial for risk management, option pricing, and portfolio allocation \cite{taylor2011asset}. While deep learning methods have shown immense potential in this domain due to their powerful nonlinear pattern recognition capabilities \cite{kim2018forecasting, liu2019novel}, their full potential remains constrained by two fundamental and often-overlooked bottlenecks in the standard modeling pipeline.

The first bottleneck is feature fidelity. Financial time series, as typical complex systems, possess an intrinsic structure that traditional statistical features cannot fully capture. The Fractal Market Hypothesis (FMH) from econophysics \cite{peters1994fractal} points out that long-range memory and multifractal properties are key dynamical characteristics of the market. This makes fractal analysis a powerful and theoretically grounded tool for our investigation. However, the application of these advanced tools faces significant limitations. First, their analytical framework is often limited to a single time series, largely failing to consider the cross-asset risk transmission mechanisms, thus ignoring the asymmetric cross-correlation phenomenon that is at the core of systemic risk \cite{longin2001extreme, ang2002asymmetric, ding2011asymmetric}. Second, the standard implementation of these methods has inherent instability, is prone to introducing spurious noise, and thus harms the fidelity of the extracted features \cite{bashan2008comparison}. These two problems together lead to a situation where the input information the model relies on is, at its source, a distorted representation of the market's true state.

The second, and more fundamental, bottleneck lies in model responsiveness. The core processing units widely adopted in deep learning models—static activation functions such as ReLU—reveal their inherent design limitations when processing highly dynamic and non-stationary financial signals. This limitation stems from a complexity mismatch \cite{wang2021chaotic}: a simple, fixed-logic processor is used to analyze an essentially chaotic and dynamic complex signal. This bottleneck means that even with perfect input features, the model's rigid internal processing mechanism cannot provide an effective dynamic response, ultimately leading to information decay and loss during internal transmission \cite{lee2019chaotic}.

In fact, these two bottlenecks are not mutually independent; they are two facets of the same underlying challenge posed by chaotic systems. In nonlinear dynamics, the long-term behavior of a chaotic system is governed by a strange attractor, which possesses an intricate fractal geometry. From this perspective, the challenge of feature fidelity is about accurately quantifying the geometric properties (i.e., the multifractal spectrum) of the market's underlying strange attractor. The challenge of model responsiveness, in turn, is about having an internal processor that can dynamically react to this geometric information. This intrinsic link between chaos and fractals provides the theoretical backbone for our proposed solution.

To systematically address these two theoretically unified challenges, this paper proposes the Fractal-Chaotic Oscillation Co-driven (FCOC) framework, a novel paradigm for volatility forecasting. The FCOC framework is built upon two synergistic innovative pillars:
\begin{itemize}
    \item A \textbf{Fractal Feature Corrector (FFC)}, designed to provide a high-fidelity market complexity metric by capturing systemic asymmetric cross-correlations and rectifying the stability deficiencies of standard fractal analysis.
    \item A \textbf{Chaotic Oscillation Component (COC)}, which replaces static activation functions with a bio-inspired dynamic system to resolve the critical complexity mismatch within the model.
\end{itemize}
The core contributions of this study can be summarized as follows:
\begin{itemize}
    \item We propose the FCOC, a new framework that systematically addresses the dual bottlenecks at both the feature level and the model level in financial forecasting.
    \item We introduce the FFC, centered on a robust OSW-MF-ADCCA implementation, which significantly improves the stability and fidelity of multifractal feature extraction.
    \item We deploy the COC, a dynamic activation system which, through a systematic exploration of the parameter space, includes two novel configurations (T9 and T10) proposed to adapt to financial dynamics.
    \item Through systematic empirical analysis, we demonstrate that the co-driven synergy between FFC and COC achieves significant performance gains over a range of benchmark models and establishes a new design philosophy for intelligent systems in complex financial environments.
\end{itemize}

The remainder of this paper is structured as follows. Section \ref{sec:related_work} reviews the related work. Section \ref{sec:methods} provides a detailed exposition of the FCOC framework's methodology. Section \ref{sec:problem_and_experiments} outlines the problem formulation and the experimental setup. Section \ref{sec:empirical_analysis} presents the empirical results and provides an in-depth discussion of the findings. Finally, Section \ref{sec:conclusion} concludes the paper.

\section{Related Work}
\label{sec:related_work}
Our research is situated at the intersection of deep learning for finance, econophysics, and computational neuroscience.

\subsection{Deep Learning Models for Financial Time Series Forecasting}
Modern financial forecasting methods have widely adopted deep learning architectures. Recurrent Neural Networks (RNNs), particularly Long Short-Term Memory (LSTM) \cite{kim2018forecasting, liu2019novel} and Gated Recurrent Units (GRU) \cite{chen2019forecasting}, are commonly used for their ability to handle temporal dependencies. In recent years, more advanced architectures such as the Transformer \cite{vaswani2017attention} and state-space models like Mamba have also been explored. For instance, Lu and Xu \cite{lu2024trnn} proposed an efficient Time-series Recurrent Neural Network (TRNN) to improve training efficiency. Chen et al. \cite{chen2024improved} developed a hybrid model combining a Temporal Convolutional Network (TCN) with BiGRU for new energy stock index forecasting. While these works have made significant progress in model architecture, they almost universally rely on traditional static activation functions, thus failing to fundamentally address the internal model responsiveness problem.

\subsection{Multifractal Analysis of Financial Time Series}
To overcome the informational limitations of raw time series, a more theoretically profound approach originates from econophysics. This discipline aims to uncover the intrinsic physical laws of the market, with the multifractal nature of financial time series being a key insight that reveals the non-uniformity of volatility across different time scales \cite{kantelhardt2002multifractal}.

In response, econophysics has developed an evolving toolkit of methods, progressing from Detrended Fluctuation Analysis (DFA) \cite{peters1994fractal} to Multifractal DFA (MF-DFA) \cite{kantelhardt2002multifractal}. A significant advance came with the recognition of market asymmetry, leading to Asymmetric MF-DFA (A-MFDFA) \cite{lee2017asymmetric, lee2018asymmetric}. However, these methods were long confined to single-asset analysis, overlooking the systemic interactions of the market as a whole. In fact, asymmetric cross-correlations between different assets are a key driver of systemic market risk, especially during downturns \cite{longin2001extreme, ang2002asymmetric, ding2011asymmetric}. Recent work has continued to validate this, with Yu et al. \cite{yu2022novel} successfully combining multifractal analysis with GRU networks and Wang and Lee \cite{wang2023stock} applying modified MF-ADCCA to forecast stock index volatility. Although subsequent research began to address cross-correlations with more complex techniques, they commonly inherited a persistent technical flaw from earlier methods: the use of non-overlapping segmentation, which has been shown to introduce spurious fluctuations and compromise the stability of fractal measurements \cite{bashan2008comparison}. Our Fractal Feature Corrector (FFC) is designed to solve both of these historical problems simultaneously. By using Multifractal Asymmetric Detrended Cross-Correlation Analysis (MF-ADCCA) \cite{cao2014detrended} as its core to capture systemic interactions, and employing a robust Overlapping Sliding Window (OSW) implementation to correct for instability \cite{tang2019research}, the FFC aims to provide a more reliable and higher-fidelity set of complexity features than previously available.

\subsection{Dynamic and Chaotic Activation Functions}
The concept of enhancing neural networks with internal dynamic components originates from computational neuroscience. Unlike standard neural networks, the biological brain is believed to operate based on continuous and chaotic neural oscillations \cite{freeman2000neurodynamics}. Inspired by this, academia has begun to explore replacing static activation functions with dynamic systems \cite{wang2021chaotic}. This direction also aligns with cutting-edge paradigms in econophysics, which posit that financial markets can be viewed as a dynamic energy field where assets behave like interacting, chaotic oscillators. From this perspective, an oscillator with excitatory and inhibitory mechanisms is not merely a computational tool but a physical analogy for the fundamental push-pull dynamics of market forces.

Therefore, our adoption of the Chaotic Oscillation Component (COC) is dually motivated: computationally, it solves the complexity mismatch problem; physically, it provides a more plausible model for the phenomena we aim to forecast. Our COC is built upon the foundational work on the Lee oscillator \cite{lee2004transient}. While generative models like the VAR-VAE proposed by Leushuis \cite{leushuis2025probabilistic} have started to incorporate probabilistic dynamics in latent spaces, our work focuses on a different, complementary goal: embedding dynamics directly into the neuron's fundamental activation process. Our core contribution in this area is not only the systematic application of this concept to financial volatility forecasting but also the engineering of two novel oscillator configurations (T9 and T10) through parameter space exploration, specifically to better capture the unique dynamics of financial markets, such as abrupt regime switching.

\section{The FCOC Framework and Methodology}
\label{sec:methods}
\begin{figure*}[!ht]
    \centering
    \includegraphics[width=\textwidth]{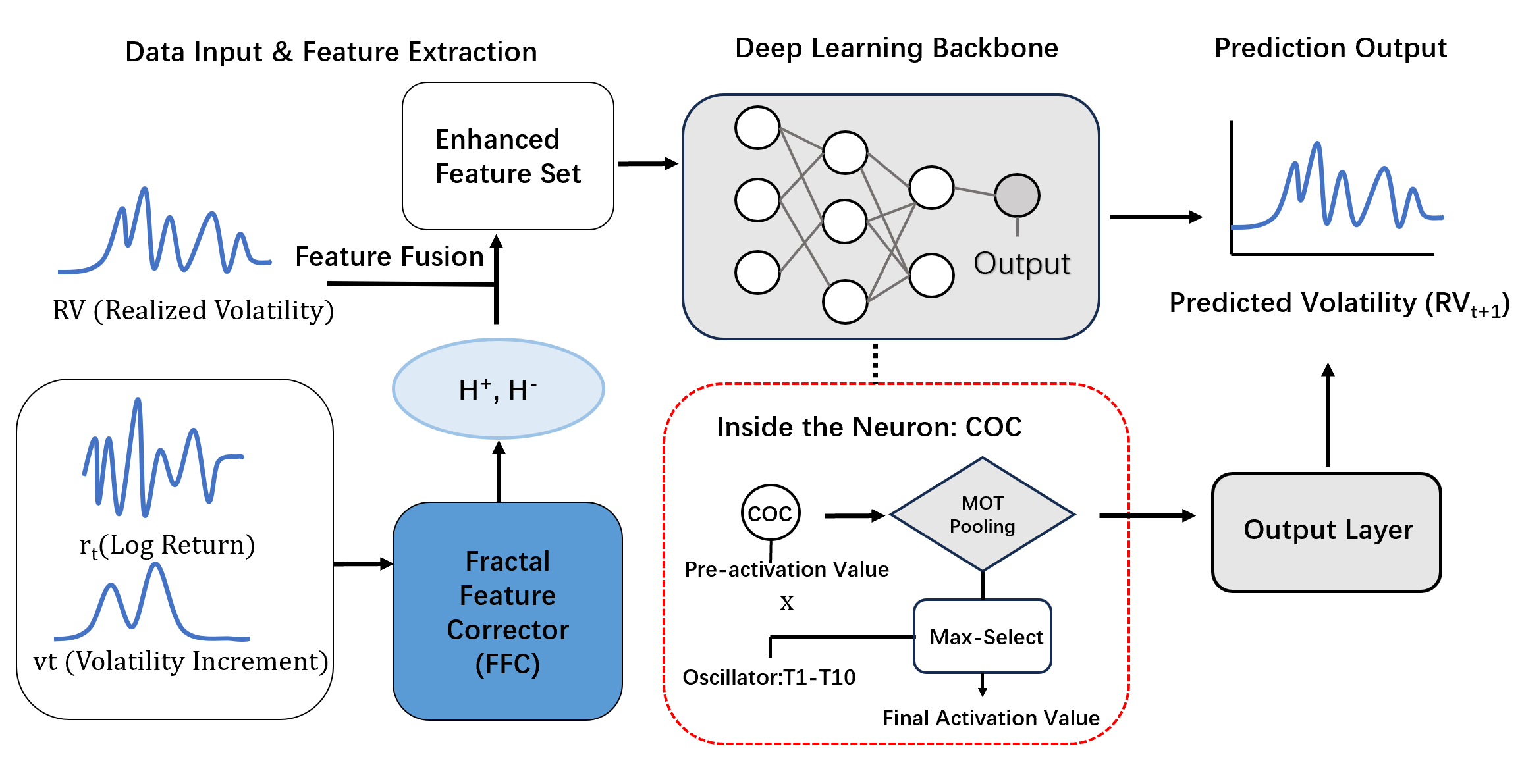}
    \caption{Conceptual Architecture of the FCOC Framework.}
    \label{fig:fcoc_flowchart}
\end{figure*}

The proposed FCOC framework addresses the dual challenges of robust feature extraction and dynamic model responsiveness. The overall architecture of the framework is depicted in Figure \ref{fig:fcoc_flowchart}. This section provides a comprehensive technical exposition of the two innovative pillars that constitute the framework. First, we detail the FFC, whose core is our novel OSW-MF-ADCCA algorithm, designed to generate high-fidelity fractal features. Second, we elaborate on the COC, which fundamentally upgrades the network's internal processing units from static activation functions to a dynamic chaotic system.

\subsection{Fractal Feature Corrector (FFC)}
\label{subsec:osw-mf-adcca}
The Fractal Feature Corrector is designed to address the feature fidelity bottleneck by resolving the two fundamental limitations of prior fractal analyses identified in our introduction: their confinement to single-asset analysis and their inherent measurement instability. At its core is our robust OSW-MF-ADCCA algorithm, a technique that quantifies the asymmetric and multifractal cross-correlations between two non-stationary time series. Its core procedure involves analyzing the cumulative deviation profiles of the series through systematically overlapping sub-intervals. For two time series, $r_x(t)$ and $r_y(t)$ (for $t=1, \dots, N$), the detailed implementation steps are as follows:

\textbf{Step 1: Profile Construction} \\
The cumulative deviation series, denoted as profiles $\mathcal{P}_x(k)$ and $\mathcal{P}_y(k)$, are first generated from the original series:
\begin{equation}
    \mathcal{P}_x(k) = \sum_{t=1}^{k} (r_x(t) - \mu_x), \quad k=1, \dots, N
\end{equation}
\begin{equation}
    \mathcal{P}_y(k) = \sum_{t=1}^{k} (r_y(t) - \mu_y), \quad k=1, \dots, N
\end{equation}
where $\mu_x$ and $\mu_y$ represent the mean values of the entire series $r_x$ and $r_y$, respectively. To facilitate the asymmetry analysis, the primary market return series, $r_x(t)$, is designated as the proxy for the local market trend.

\textbf{Step 2: Overlapping Segmentation} \\
The profiles are partitioned into segments of length $s$ using a sliding window. The window advances with a stride of $s_{step}$, which is governed by an overlap ratio $\rho$ ($0 \le \rho < 1$):
\begin{equation}
    s_{step} = \lfloor s \cdot (1 - \rho) + 0.5 \rfloor
\end{equation}
This process yields $N_{seg} = \lfloor (N - s) / s_{step} \rfloor + 1$ overlapping segments. In this work, an overlap ratio of $\rho=1/3$ is used, a choice that strikes a balance between enhancing statistical stability and maintaining computational efficiency, consistent with practices in related studies \cite{tang2019research}.

\textbf{Step 3: Local Detrending and Trend Discrimination} \\
For each segment $j$ ($j=1, \dots, N_{seg}$), a polynomial of order $m=2$ is fitted to remove the local trend. The choice of a quadratic polynomial ($m=2$) offers a robust balance, effectively removing complex non-linear local trends common in financial series without overfitting to short-term noise. Let $p_x^{(j)}(i)$ and $p_y^{(j)}(i)$ be the polynomial fits for the segment. The local detrended fluctuation, $F^2_{seg}(s, j)$, is calculated as:
\begin{equation}
    F^2_{seg}(s, j) = \frac{1}{s} \sum_{i=1}^{s} |(\mathcal{P}_x(i) - p_x^{(j)}(i)) \cdot (\mathcal{P}_y(i) - p_y^{(j)}(i))|
\end{equation}
Within the same segment, the slope of a linear fit to the index proxy series, denoted as $\beta_j$, is used to identify the trend's direction. A positive trend corresponds to $\beta_j > 0$, and a negative trend otherwise.

\textbf{Step 4: Directional $q$-order Fluctuation Functions} \\
Fluctuation functions are then computed by averaging over segments with positive and negative trends separately. For a given order $q$, these directional functions are defined as:
\begin{equation}
    F_{q}^{+}(s) = \left[ \frac{1}{N_{pos}} \sum_{j=1}^{N_{seg}} \frac{1+\text{sgn}(\beta_j)}{2} [F^2_{seg}(s,j)]^{q/2} \right]^{1/q}
\end{equation}
\begin{equation}
    F_{q}^{-}(s) = \left[ \frac{1}{N_{neg}} \sum_{j=1}^{N_{seg}} \frac{1-\text{sgn}(\beta_j)}{2} [F^2_{seg}(s,j)]^{q/2} \right]^{1/q}
\end{equation}
where $N_{pos}$ and $N_{neg}$ are the total counts of segments with positive and negative trends, respectively. The separation is achieved using the sign function, $\text{sgn}(\beta_j)$, which returns $+1$ for a positive local trend slope ($\beta_j > 0$), $-1$ for a negative slope, and $0$ otherwise. This makes the term $(1+\text{sgn}(\beta_j))/2$ an indicator that equals $1$ only for positive-trend segments, while $(1-\text{sgn}(\beta_j))/2$ acts as an indicator for negative-trend segments. For the special case of $q=0$, the averaging is performed in the logarithmic domain to avoid singularities. The fluctuation function is calculated as:
\begin{equation}
    F_{0}(s) = \exp \left( \frac{1}{2N_{seg}} \sum_{j=1}^{N_{seg}} \ln [F^2_{seg}(s,j)] \right)
\end{equation}
with analogous definitions for the directional cases $F_{0}^{+}(s)$ and $F_{0}^{-}(s)$ by applying the respective indicator functions within the summation.

\textbf{Step 5: Estimation of Generalized Hurst Exponents} \\
The existence of long-range power-law cross-correlations is indicated if the fluctuation functions scale with the segment size $s$ as follows:
\begin{equation}
    F_q(s) \propto s^{H(q)}; \quad F_q^+(s) \propto s^{H^+(q)}; \quad F_q^-(s) \propto s^{H^-(q)}
\end{equation}
where the exponents $H(q)$, $H^+(q)$, and $H^-(q)$ are the generalized Hurst exponents. They are determined from the slope of a log-log plot of the fluctuation function versus segment size $s$. Our analysis focuses on the case of $q=2$. This choice is standard in the econophysics literature for analyzing volatility persistence, as it directly relates to the second moment (variance) of the fluctuations and provides a measure analogous to the classical Hurst exponent for long-range correlations \cite{kantelhardt2002multifractal, podobnik2008detrended}. Here, $H(2) > 0.5$ suggests persistent cross-correlations and $H(2) < 0.5$ suggests anti-persistent cross-correlations.

\textbf{Step 6: Rolling Window Feature Generation} \\
To capture the temporal dynamics of market correlations, the Hurst exponents are calculated not once, but continuously over time. A rolling-window approach is implemented, where the exponent is calculated for each day using data from the preceding $T$ days. This methodology generates time-varying feature vectors. The entire procedure is formalized in Algorithm \ref{alg:rolling_osw_mfadcca}.

\begin{algorithm}[h!]
\caption{Rolling Window Feature Generation via OSW-MF-ADCCA}
\label{alg:rolling_osw_mfadcca}
\begin{algorithmic}[1]
\Require
    \State Primary time series $r_X$ of length $N$
    \State Secondary time series $r_Y$ of length $N$
    \State Rolling window size $T$
    \State Step size for window sliding $k$
    \State Overlap ratio $\rho$
\Ensure
    \State Hurst exponent series $\mathcal{H}_{overall}, \mathcal{H}_{positive}, \mathcal{H}_{negative}$

\Statex 
\Function{Generate\_Hurst\_Features}{$r_X, r_Y, T, k, \rho$}
    \State Initialize $\mathcal{H}_{all}, \mathcal{H}_{pos}, \mathcal{H}_{neg} \gets \text{empty lists}$
    \State $N_w \gets \lfloor (N - T) / k \rfloor + 1$ \Comment{Number of windows}
    \For{$i = 0$ \textbf{to} $N_w - 1$}
        \State $start \gets i \times k$
        \State $end \gets start + T$
        \State $s_X \gets r_X[start \dots end]$
        \State $s_Y \gets r_Y[start \dots end]$
        
        \State \Comment{Calculate exponents for the sub-window}
        \State $H, H^{+}, H^{-} \gets \text{Calc\_Hurst}(s_X, s_Y, q=2, \rho)$
        
        \State Append $H$ to $\mathcal{H}_{all}$
        \State Append $H^{+}$ to $\mathcal{H}_{pos}$
        \State Append $H^{-}$ to $\mathcal{H}_{neg}$
    \EndFor
    \State \textbf{return} $\mathcal{H}_{all}, \mathcal{H}_{pos}, \mathcal{H}_{neg}$
\EndFunction
\end{algorithmic}
\end{algorithm}

\subsection{Chaotic Oscillation Component (COC)}
\label{subsec:coc}
The second pillar of the FCOC framework, the Chaotic Oscillation Component, addresses a fundamental limitation within deep learning models: the dynamic inertness of conventional activation functions. To overcome this deficiency, we fundamentally upgrade the activation function to a dynamic chaotic micro-system. The process involves two primary stages: first, distilling the complex behavior of multiple oscillators into a library of candidate functions, and second, adaptively selecting from this library to produce a final activation value.

\subsubsection{Core Engine: Lee Oscillator with Retrograde Signaling (LORS)}
The decision to replace a static activation function with a dynamic oscillator represents a fundamental paradigm shift grounded in both econophysics and neuroscience. Classical neural networks, with their simple neuron models, have been criticized for being far simpler than their biological counterparts. Modern neuroscience reveals that the brain does not operate on a simple feed-forward firing of static units; rather, it functions on a substrate of continuous and chaotic oscillations known as brainwaves \cite{freeman2000neurodynamics}. It is this underlying oscillatory dynamic that gives rise to complex cognitive functions like memory and perception. Inspired by this, we consider that financial markets, as complex adaptive systems driven by collective human behavior, are more faithfully modeled as a field of interacting oscillators than as a system mapped by static functions.

\begin{figure}[h!]
    \centering
    \includegraphics[width=0.5\columnwidth]{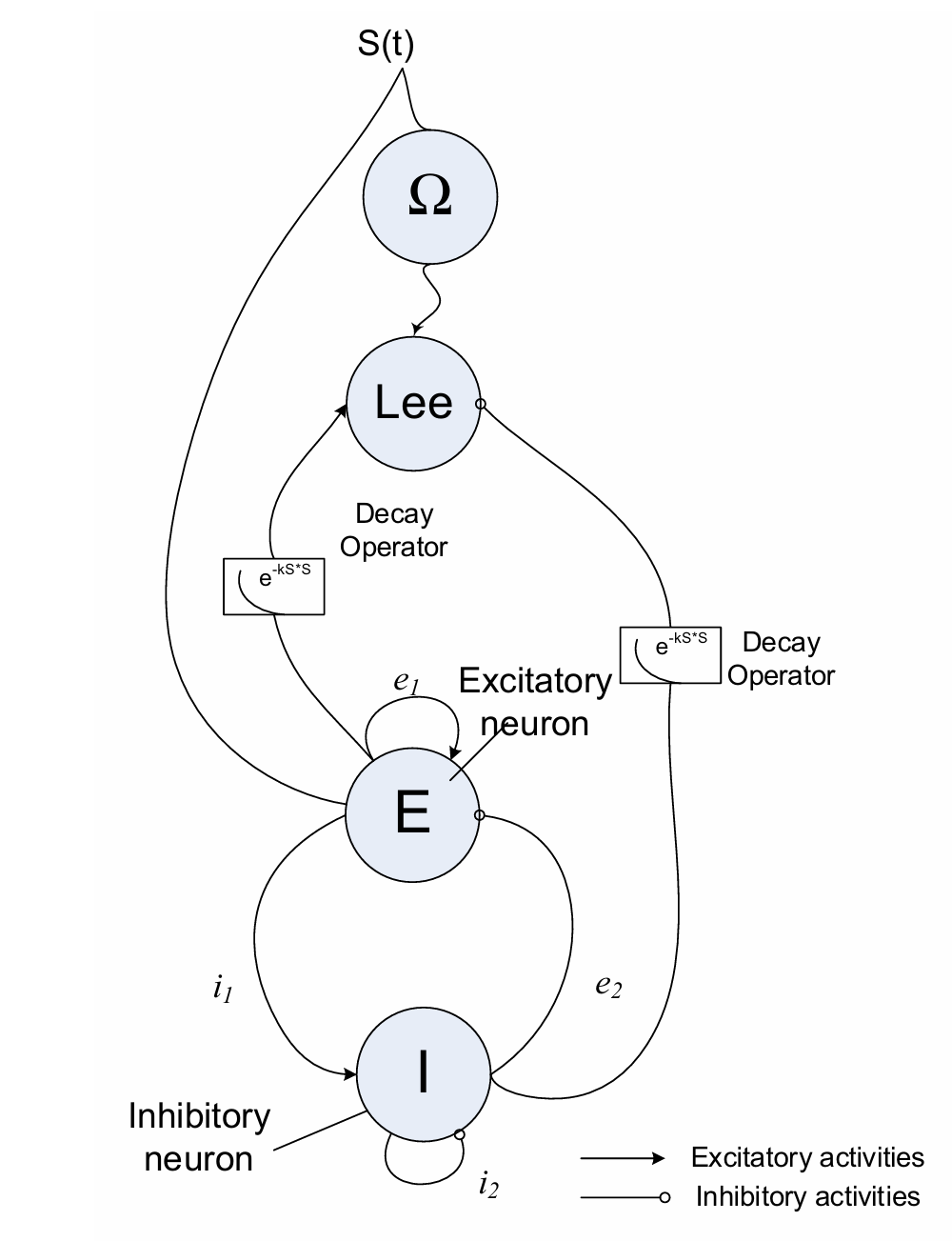}
    \caption{Neural architecture of the Lee Oscillator with Retrograde Signaling (LORS).}
    \label{fig:lors_architecture}
\end{figure}

The Lee oscillator, with its design rooted in emulating biologically plausible neural dynamics such as Progressive Memory Recall \cite{lee2006progressive}, provides the ideal computational primitive for this paradigm. It allows a system to perform gradual feedback and self-correction when faced with incomplete or noisy inputs through transient chaotic behavior, a mechanism strikingly similar to how market participants adapt to new information. Specifically, this study employs an advanced variant: the Lee Oscillator with Retrograde Signaling. This model enhances the original design \cite{lee2004transient} by incorporating retrograde signaling mechanisms observed in neuroscience \cite{levitan2002neuron, wong2008wind}, further boosting its biological plausibility. The neural architecture of the LORS is depicted in Figure \ref{fig:lors_architecture}.

Its dynamic behavior is strictly governed by the following set of equations:
\begin{align}
    f(\mu; x) &= \tanh(\mu x) \\
    E_{t+1} &= f(a_1 \text{LORS}_{t} + a_2 E_{t} - a_3 I_{t} + a_4 S_{t} - \xi_E) \\
    I_{t+1} &= f(b_1 \text{LORS}_{t} - b_2 E_{t} - b_3 I_{t} + b_4 S_{t} - \xi_I) \\
    S_{t} &= i + e \cdot \tanh(i) \\
    \Omega_{t+1} &= f(S_{t}) \\
    \text{LORS}_{t} &= [E_{t} - I_{t}] \cdot e^{-k S^2_{t}} + \Omega_{t}
\end{align}
where $E_t$ and $I_t$ represent the states of the excitatory and inhibitory neurons, respectively. The term $i$ is the base input stimulus (i.e., the pre-activation value from the neural network layer), which is modulated into the external stimulus $S_t$ via the ratio $e$. The term $\Omega_t$ represents the retrograde signal itself. The parameters $a_i$ and $b_i$ are weights governing the internal connections, while $\xi_E$ and $\xi_I$ are the corresponding threshold biases. $k$ is an attenuation factor, $\mu$ is a gain parameter, and $\text{LORS}_t$ is the oscillator's final output at time step $t$. The hyperbolic tangent ($\tanh$) is used as the base non-linearity $f(\mu; x)$ due to its bounded and sigmoidal nature, which is well-suited for modeling neural firing rates.

A key aspect of our work is the utilization of a diverse set of ten \textit{parameterized} Lee oscillators. The initial eight types, rigorously derived from systematic studies of the oscillator's bifurcation behavior \cite{lee2019chaotic}, already provide a broad range of dynamics. These include simple bifurcations (e.g., T1, T4), dense chaotic regions (T2, T3), and more complex structures with periodic windows (T5, T7). To better capture the specific topological features inherent in financial market dynamics, we extend this set through a systematic exploration of the parameter space. This leads to the design of two additional configurations, T9 and T10, which were empirically identified through parameter space exploration to better capture the unique dynamics of financial markets.

\begin{figure}[h!]
    \centering
    \includegraphics[width=0.5\columnwidth]{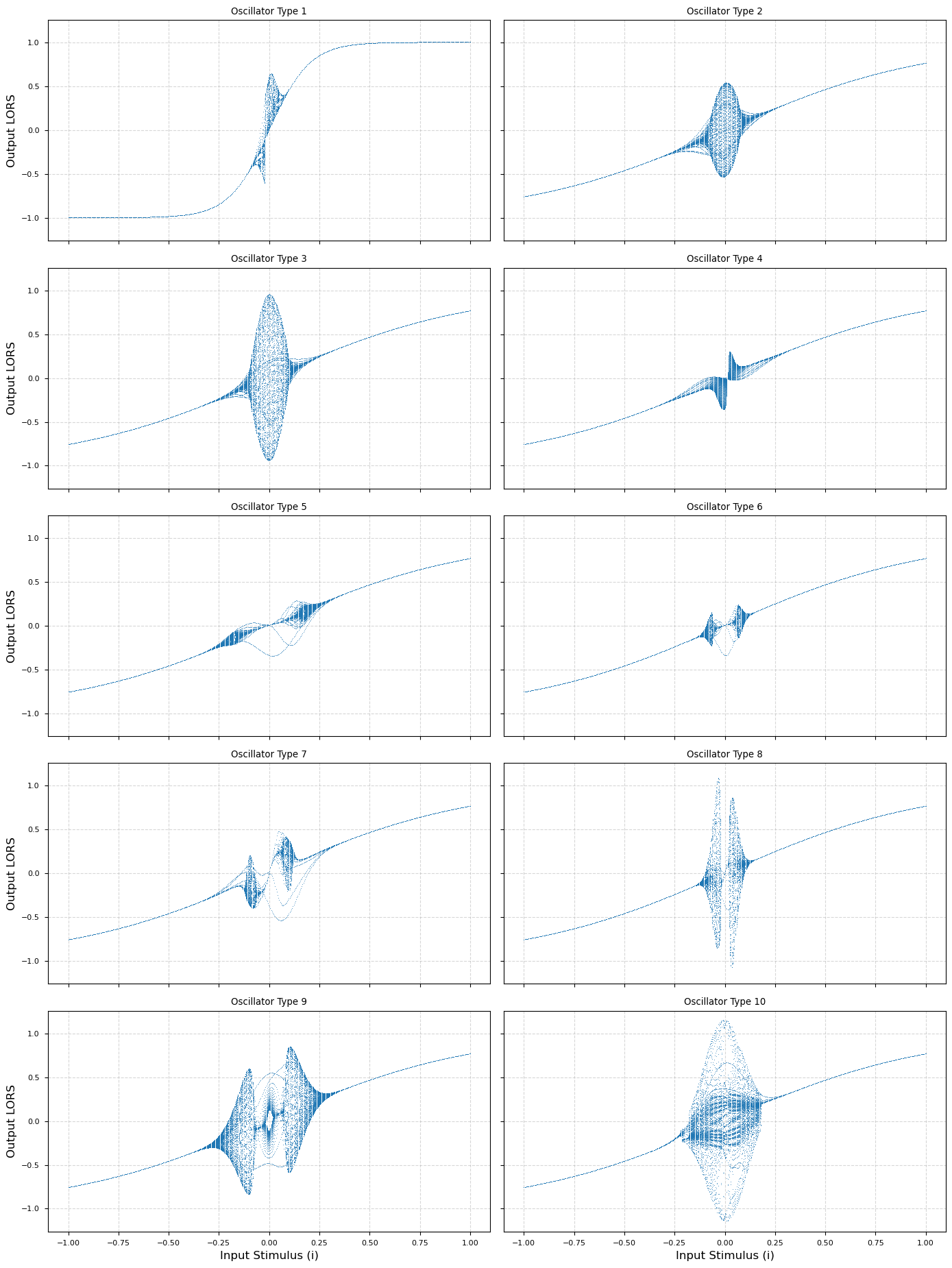}
    \caption{Bifurcation diagrams for the ten LORS types, showcasing the diverse dynamic behaviors.}
    \label{fig:bifurcation_diagrams}
\end{figure}

The most notable feature of T9 is its multi-modal and highly complex structure around the center, which appears not as a single chaotic cloud but as multiple distinct "lobes" or sub-regimes (see Figure \ref{fig:bifurcation_diagrams}). This is specifically designed to be analogous to a market that not only exhibits volatility but can abruptly switch between different types of volatile behavior. Conversely, the main characteristic of T10 is its extremely wide and dense chaotic region around the zero-input stimulus. Compared to T2 or T3, it represents a state of generalized and persistent uncertainty, where a much wider range of small input perturbations results in highly unpredictable outcomes.

The inclusion of these two targeted configurations enriches the COC's library, providing a more comprehensive set of dynamic responses to better model the varied and shifting states of financial volatility. The specific parameter settings are detailed in Table \ref{tab:lee_params}, and their corresponding bifurcation patterns are visually illustrated in Figure \ref{fig:bifurcation_diagrams}.

\begin{table}[h!]
\centering
\caption{Parameter settings for the 10 types of Lee Oscillators used in experiments.}
\label{tab:lee_params}
\resizebox{\linewidth}{!}{%
\begin{tabular}{l|cccccccccc}
\toprule
 & \textbf{T1} & \textbf{T2} & \textbf{T3} & \textbf{T4} & \textbf{T5} & \textbf{T6} & \textbf{T7} & \textbf{T8} & \textbf{T9} & \textbf{T10} \\
\midrule
$a_1$ & 0.0 & 0.5 & 0.5 & -0.5 & -0.9 & -0.9 & -5.0 & -5.0 & 1.0 & 3.0 \\
$a_2$ & 5.0 & 0.55 & 0.6 & 0.55 & 0.9 & 0.9 & 5.0 & 5.0 & -1.0 & 3.0 \\
$a_3$ & 5.0 & 0.55 & 0.55 & 0.55 & 0.9 & 0.9 & 5.0 & 5.0 & -1.0 & 3.0 \\
$a_4$ & 1.0 & -0.5 & 0.5 & -0.5 & -0.9 & -0.9 & -5.0 & -5.0 & -1.0 & 2.0 \\
\midrule
$b_1$ & 0.0 & 0.5 & -0.5 & -0.5 & 0.9 & 0.9 & 1.0 & 1.0 & -1.0 & 0.45 \\
$b_2$ & -1.0 & -0.55& -0.6 & -0.55& -0.9 & -0.9 & -1.0 & -1.0 & 2.0 & -0.45 \\
$b_3$ & 1.0 & -0.55& -0.55& -0.55& -0.9 & -0.9 & -1.0 & -1.0 & 2.0 & -0.45 \\
$b_4$ & 0.0 & -0.5 & 0.5 & 0.5 & 0.9 & 0.9 & 1.0 & 1.0 & -1.0 & 1.0 \\
\midrule
$\mu$ & 5 & 1 & 1 & 1 & 1 & 1 & 1 & 1 & 1 & 1 \\
$k$ & 500 & 50 & 50 & 50 & 50 & 300 & 50 & 300 & 50 & 50 \\
$e$ & 0.001& 0.001& 0.001& 0.001& 0.001& 0.001& 0.001& 0.001& 0.001& 0.001\\
\bottomrule
\end{tabular}%
}
\end{table}

\subsubsection{Stage 1: Distillation into a Meta-Activation Library}
While the LORS provides a wealth of dynamic information, its 100-step temporal trajectory output is architecturally incompatible with standard neural network layers. To bridge this gap, we design a critical processing stage: Max-over-Time (MoT) Pooling. This process is not a mere dimensionality reduction but a fundamental distillation. It transforms the entire dynamic trajectory of an oscillator into a single and salient scalar value, effectively creating a unique, static, yet highly nonlinear meta-activation function.

\begin{figure}[h!]
    \centering
    \includegraphics[width=0.5\columnwidth]{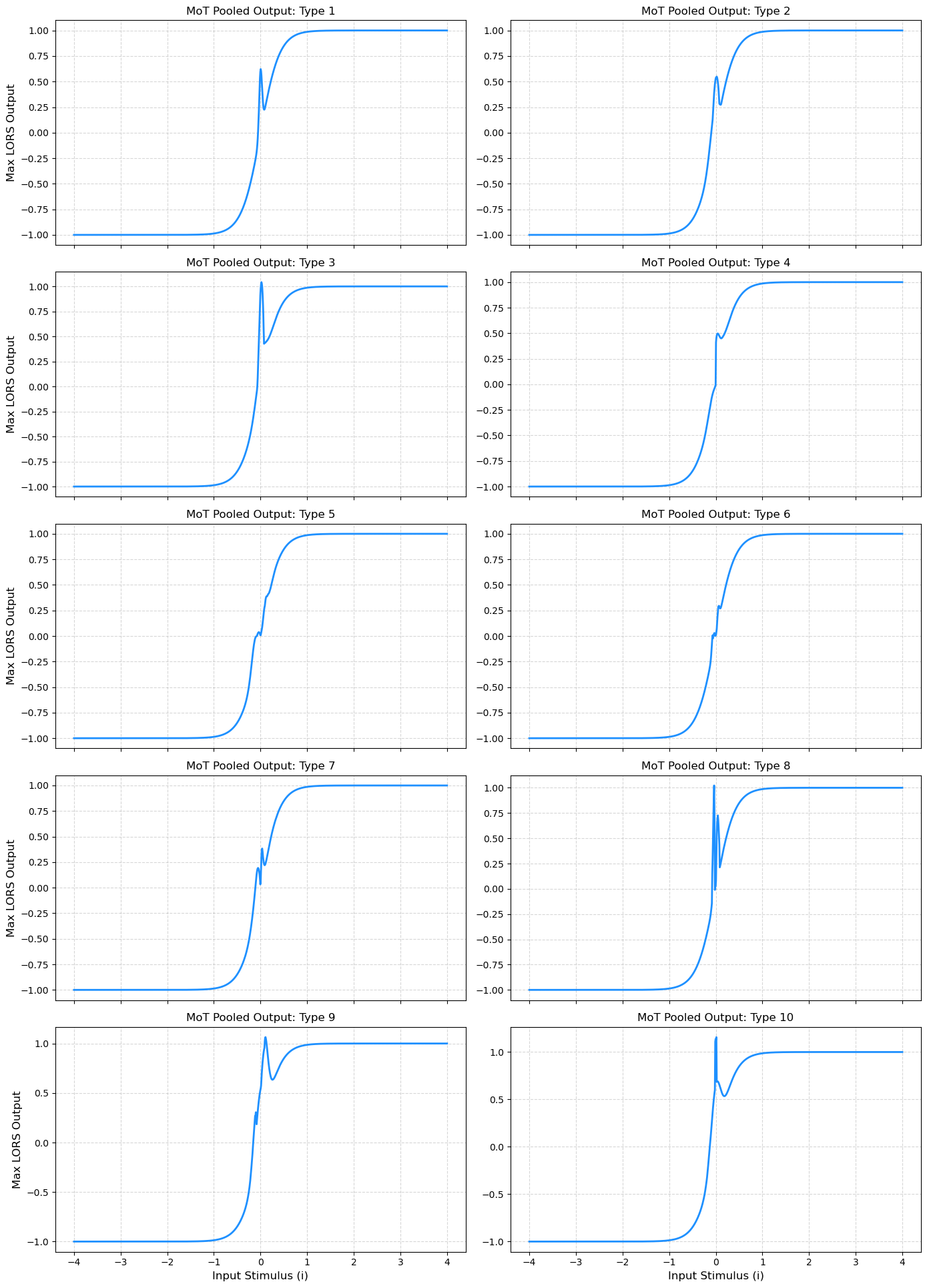}
    \caption{The ten distinct meta-activation functions, $f_{\text{type}}(x)$, generated by applying Max-over-Time (MoT) pooling.}
    \label{fig:meta_activations}
\end{figure}

The procedure, formalized in Algorithm \ref{alg:mot_pooling}, is applied to each of the ten oscillator types for a given pre-activation value $x$. The result is a library of ten distinct meta-activation functions, $\{f_{\text{T1}}(x), \dots, f_{\text{T10}}(x)\}$, as visualized in Figure \ref{fig:meta_activations}. This library forms the foundation for the subsequent adaptive selection stage.

\begin{algorithm}[h!]
\caption{Meta-Activation Generation via Max-over-Time (MoT) Pooling}
\label{alg:mot_pooling}
\begin{algorithmic}[1]
\Require
    \State Pre-activation value $x \in \mathbb{R}$
    \State Set of $M=10$ Lee oscillator parameter configurations, $\mathcal{C}$
    \State Number of oscillator steps, $N=100$
\Ensure
    \State A vector of $M$ scalar meta-activation values $\mathbf{A}(x)$

\Statex
\Function{GenerateMetaActivations}{$x, \mathcal{C}$}
    \State Initialize $\mathbf{A}(x) \gets \text{empty vector of size } M$
    \For{$i = 1$ \textbf{to} $M$}
        \State $C_{\text{type}} \gets \mathcal{C}[i]$
        \State $\text{LORS}_{\text{raw\_traj}} \gets \text{Run\_Oscillator}(x, C_{\text{type}}, \text{steps}=N)$
        \State $\text{LORS}_{\text{dynamics}} \gets \text{LORS}_{\text{raw\_traj}}[1:]$
        \State $f_{\text{type}}(x) \gets \max(\text{LORS}_{\text{dynamics}})$
        \State $\mathbf{A}(x)[i] \gets f_{\text{type}}(x)$
    \EndFor
    \State \textbf{return} $\mathbf{A}(x)$
\EndFunction
\end{algorithmic}
\end{algorithm}

\subsubsection{Stage 2: Final Activation via Maximum Response Selection}
\label{subsubsec:selection_stage}
Once the library of ten meta-activation functions is generated for a given input $x$ (producing the vector $\mathbf{A}(x)$ as per Algorithm \ref{alg:mot_pooling}), the final step is to produce a single activation output. For this, we adopt the Maximum Response Selection (Max-Select) strategy. This choice is not arbitrary but is deliberately grounded in the Winner-Takes-All (WTA) principle, a computational mechanism with deep roots in theoretical neuroscience and continued relevance in modern cognitive science. The WTA concept, foundational to the theory of efficient neural coding \cite{barlow1972single}, posits that from a group of competing processing units, only the one with the strongest response should determine the final output. This principle is considered a cornerstone of models explaining decision-making and perception \cite{heathcote2022winner}.

The theoretical justification for using Max-Select is twofold. First, from a neuro-computational perspective, each of our ten oscillators represents a distinct potential dynamic regime (e.g., stable, bifurcating, wide-band chaotic). For any given input stimulus $x$, it is plausible that one specific dynamic regime is predominantly responsible for the system's subsequent behavior. The Max-Select strategy, by implementing a hard form of WTA, identifies and propagates the response from this single, most dominant regime. An alternative like averaging would dilute this salient signal by mixing it with weaker, less relevant dynamic responses, thereby obscuring the critical information.

Second, from a modeling perspective, Max-Select offers a robust, parameter-free alternative to more complex, learnable mechanisms like an attention layer. While an attention mechanism could learn to weigh the oscillators, it would introduce additional parameters and computational overhead, increasing the risk of overfitting. In contrast, our principled, biologically-inspired approach is both efficient and highly effective, as confirmed by our extensive empirical validation.
\begin{equation}
    f_{\text{Lee}}(x) = \max \left( \mathbf{A}(x) \right) = \max \left( f_{\text{T1}}(x), \dots, f_{\text{T10}}(x) \right)
\end{equation}
This approach allows us to leverage a diverse set of dynamic behaviors through a decisive selection mechanism rather than a blended compromise.

\section{Problem Formulation and Experiments}
\label{sec:problem_and_experiments}
This section formally defines the volatility forecasting problem and details the empirical validation of the FCOC framework. We begin by describing the dataset and feature generation process, and subsequently outline the complete experimental setup, including model configurations, evaluation metrics, and training protocols.

\subsection{Data Description and Feature Generation}
\label{subsec:data_description}

In this study, we use the 5-minute intraday returns and daily log-returns for two major U.S. stock market indices: the Standard \& Poor's (S\&P) 500 Index and the Dow Jones Index (DJI). The data are obtained from
the Wind Economics Database of China. The sample data consists of calculated RV data for the S\&P 500 and the DJI, spanning from December 13, 2005 to February 7, 2025, and from October 9, 2009 to February 7, 2025, respectively, which reflects the availability of the sampled high-frequency data for both indices. This combined time frame allows for a comprehensive analysis covering a wide range of market conditions, including the aftermath of the 2008 financial crisis, the subsequent period of quantitative easing, and the 2020 COVID-19 crash. 

We employ the daily realized volatility (RV) as the proxy of the true latent volatility, as it is nearly unbiased and efficient \cite{andersen2003modeling}. For a given trading day $t$ with $M$ intraday returns, RV is formally defined as:
\begin{equation}
    RV_t = \sum_{j=1}^{M} r_{t,j}^2
\end{equation}
where $r_{t,j}$ is the $j$-th intraday log return in percentage on day $t$. The use of high-frequency data to construct RV offers a significant advantage over traditional low-frequency estimators (e.g., squared daily returns), as it provides a much more precise and less noisy measure of daily price variation. In this analysis, our primary objective is to forecast the one-day-ahead RV for both indices. 

To generate the fractal features via OSW-MF-ADCCA for each index, we construct two primary input series from the high-frequency data. The first is the daily log return ($r_t$) in percentage:
\begin{equation}
    r_t = 100 \times (\ln(P_t) - \ln(P_{t-1}))
\end{equation}
where $P_t$ is the closing price on day $t$. 

The second input series is the volatility increment ($v_t$), which captures the dynamics of volatility changes. It is constructed using the realized bipower variation ($BPV_t$), a measure known for its robustness to price jumps \cite{barndorff2004power}. For a day $t$ with $M$ intraday returns $r_{t,j}$, BPV is defined as:
\begin{equation}
    BPV_t = \mu_1^{-2} \sum_{j=2}^{M} |r_{t,j}| |r_{t,j-1}|
\end{equation}
where $\mu_1 = \sqrt{2/\pi}$. We then define the volatility increment $v_t$ as the log-difference of the square root of BPV:
\begin{equation}
    v_t = \ln(\sqrt{BPV_t}) - \ln(\sqrt{BPV_{t-1}})
\end{equation}
   
The distinction between using a BPV-derived measure for feature engineering and RV as the forecast target is a deliberate methodological choice. While BPV and RV are closely related, BPV's jump-robust nature allows our fractal analysis to capture the underlying persistence of the continuous component of volatility, providing a cleaner and more stable signal of the market's memory state. Our forecast target, however, is the RV, as it represents the complete price variation, including jumps, and is therefore of greater practical and economic importance for risk management and option pricing. This approach avoids trivializing the forecasting task while leveraging the best possible signal for feature extraction.

\begin{figure*}[!ht]
    \centering
    \begin{subfigure}[b]{0.48\textwidth}
        \includegraphics[width=0.8\textwidth]{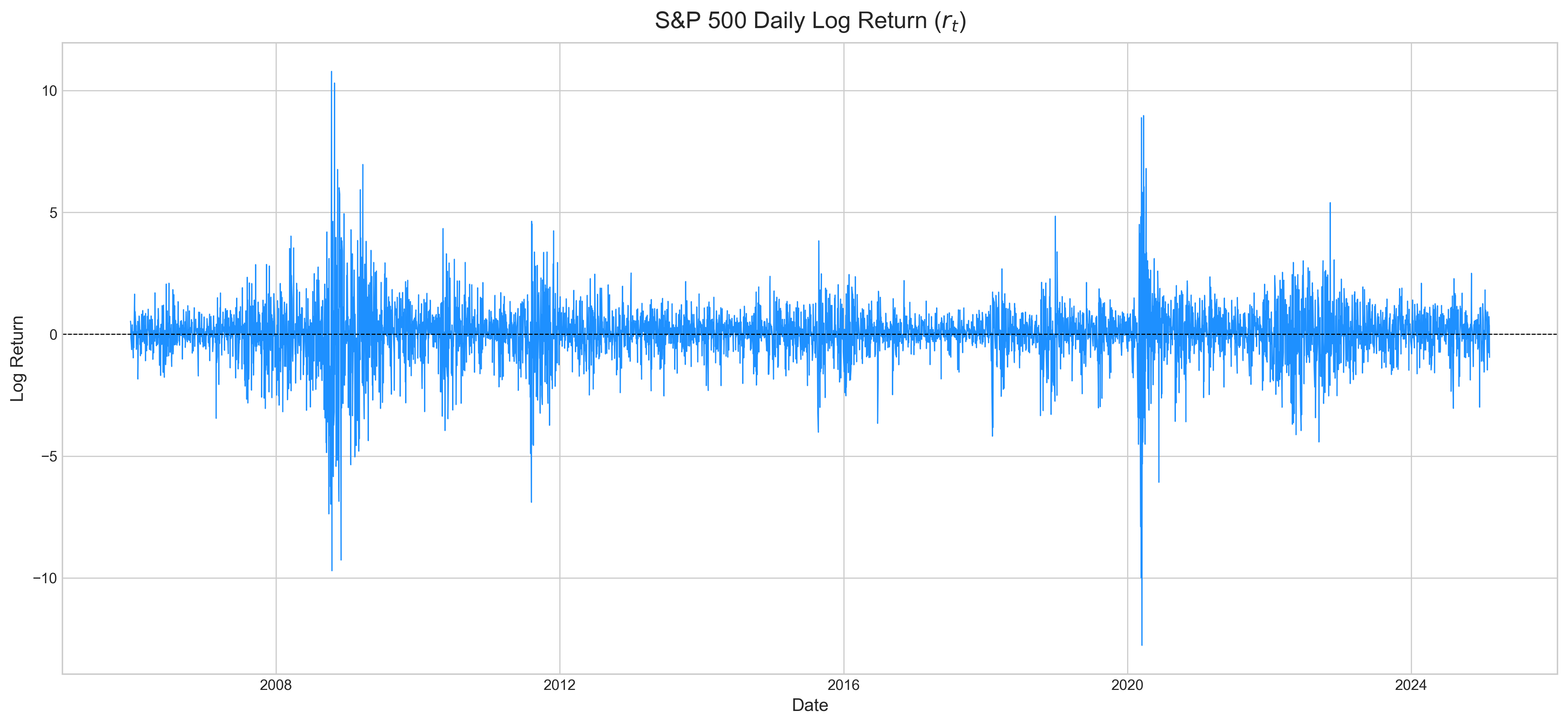}
        \caption{S\&P 500 Daily Log Return ($r_t$).}
    \end{subfigure}
    \hfill
    \begin{subfigure}[b]{0.48\textwidth}
        \includegraphics[width=0.8\textwidth]{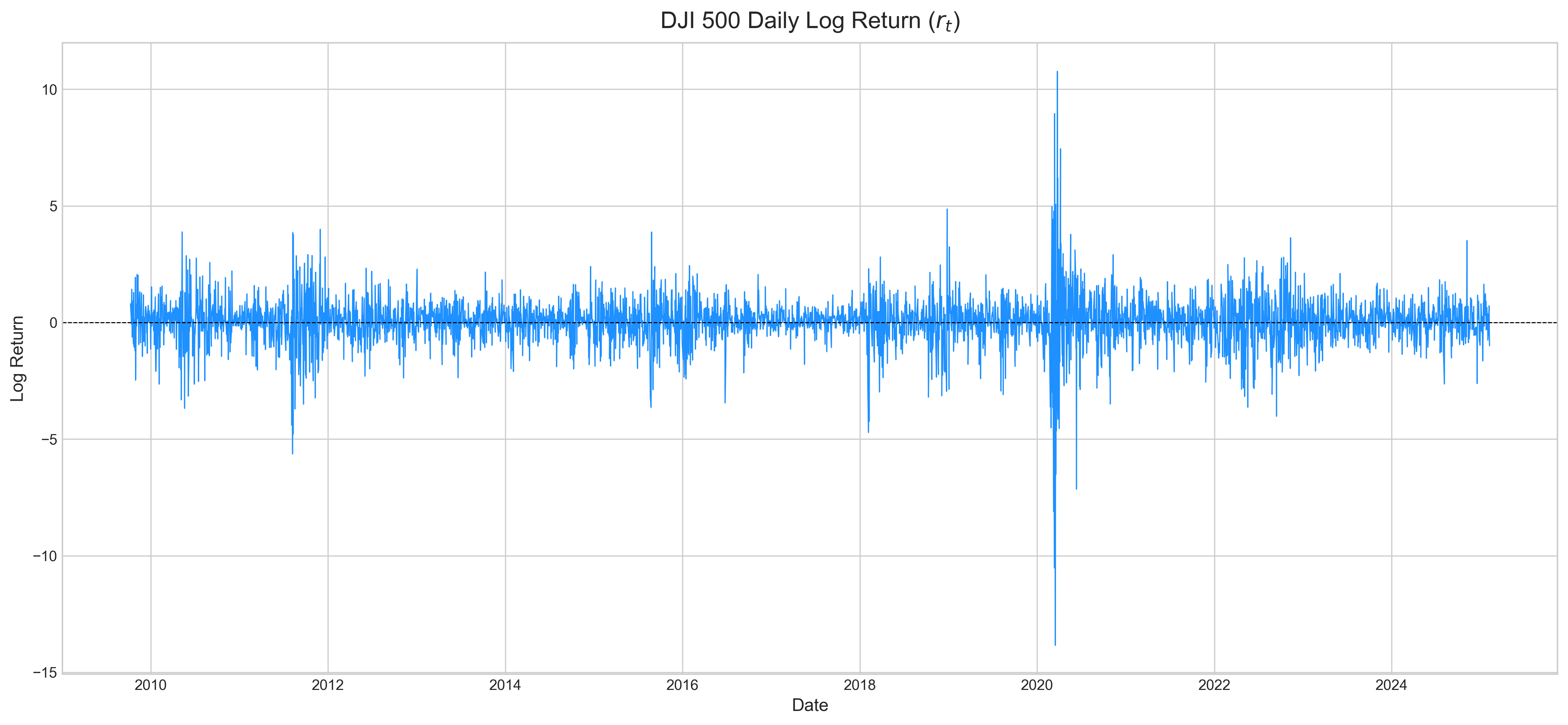}
        \caption{DJI Daily Log Return ($r_t$).}
    \end{subfigure}
    \vskip\baselineskip
    \begin{subfigure}[b]{0.48\textwidth}
        \includegraphics[width=0.8\textwidth]{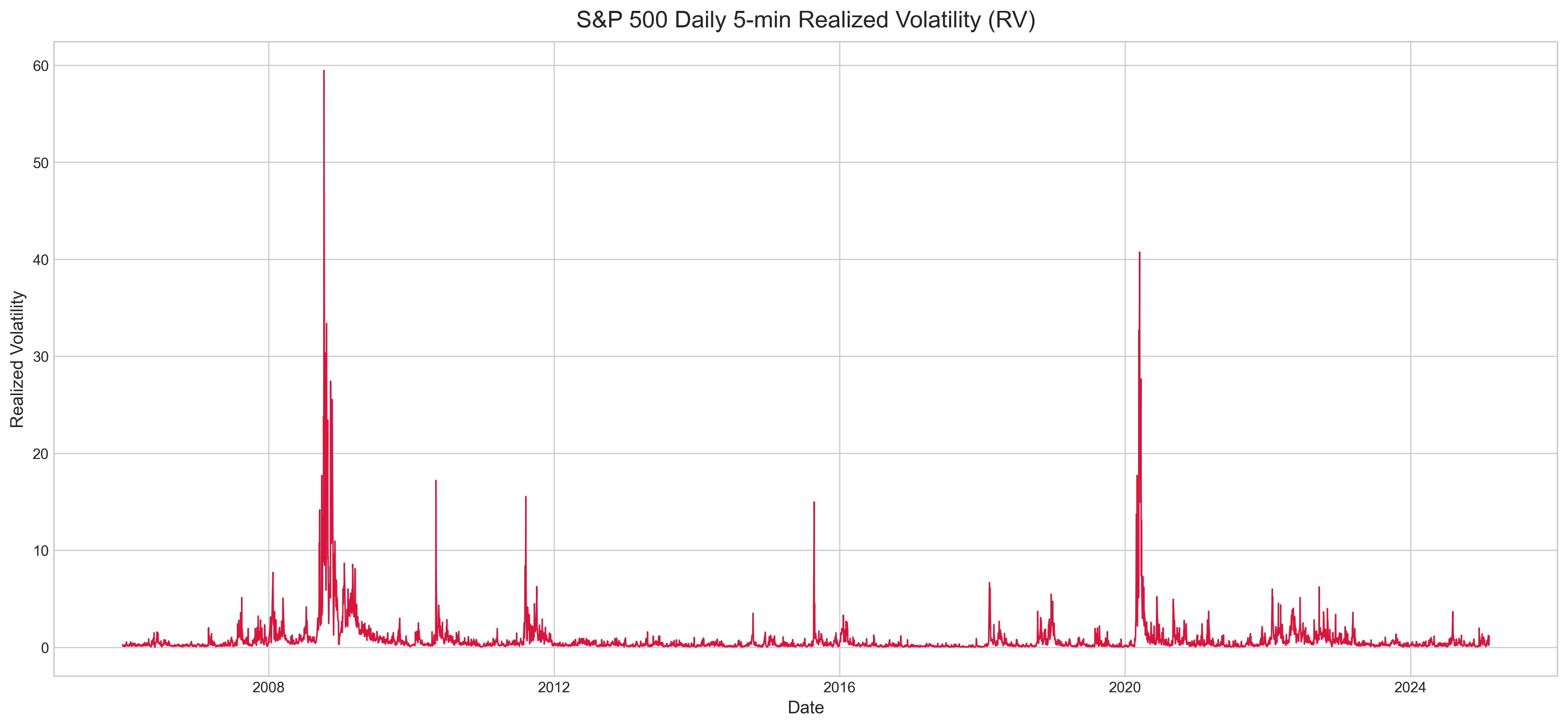}
        \caption{S\&P 500 Realized Volatility (RV).}
    \end{subfigure}
    \hfill
    \begin{subfigure}[b]{0.48\textwidth}
        \includegraphics[width=0.8\textwidth]{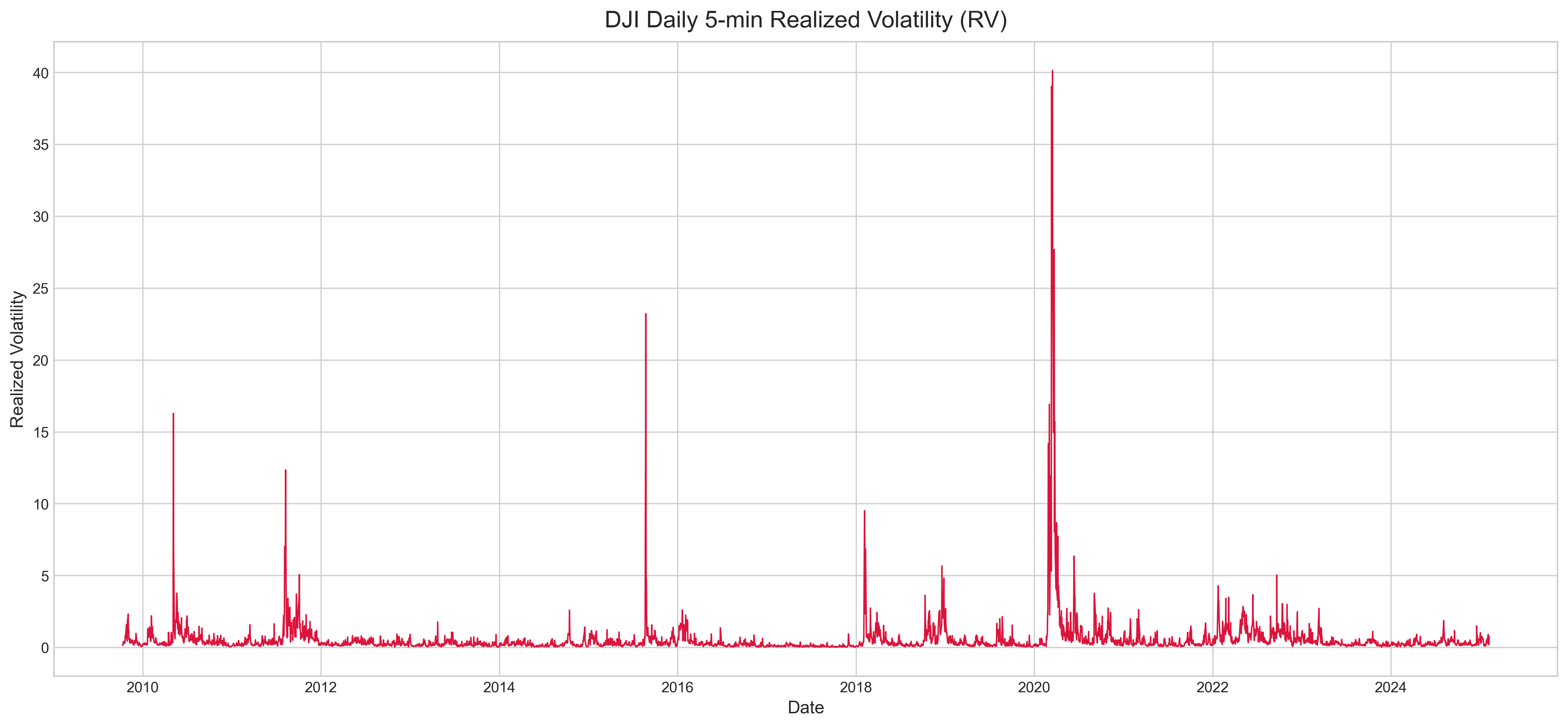}
        \caption{DJI Realized Volatility (RV).}
    \end{subfigure}
    \caption{Comparison of Log Return and Realized Volatility series for the S\&P 500 and DJI.}
    \label{fig:data_plots_combined}
\end{figure*}

The characteristics of these key time series are presented visually in Figure \ref{fig:data_plots_combined} and statistically in Table \ref{tab:descriptive_stats_combined}. The plots visually confirm the rationale for focusing on RV; while log returns appear highly noisy, the RV series for both indices exhibit clear structural patterns like volatility clustering, making them more statistically tractable targets. The descriptive statistics further confirm the stylized facts of financial data. For both indices, the RV and $r_t$ series exhibit extremely high kurtosis and significant Jarque-Bera (JB) statistics, rejecting normality and indicating fat-tailed distributions. Furthermore, the Augmented Dickey-Fuller (ADF) and KPSS tests confirm that the input series ($r_t$ and $v_t$) are stationary, validating their direct use in our models. These properties motivate the use of advanced architectures for capturing such complex dynamics.

\begin{table*}[!ht]
\centering
\caption{Descriptive statistics for key time series of the S\&P 500 and DJI}
\label{tab:descriptive_stats_combined}
\resizebox{\textwidth}{!}{%
\begin{threeparttable}
\begin{tabular}{llcccccccc}
\toprule
\textbf{Index} & \textbf{Variable} & \textbf{Mean} & \textbf{Max} & \textbf{Min} & \textbf{Std. Dev.} & \textbf{Kurtosis} & \textbf{JB} & \textbf{ADF} & \textbf{KPSS} \\
\midrule
\multirow{3}{*}{S\&P 500} & RV & 0.8538 & 59.4474 & 0.0088 & 2.2696 & 175.02 & 5903183.63*** & -7.326*** & 0.5817 \\
& $r_t$ & 0.0317 & 10.7811 & -12.7652 & 1.2453 & 12.01 & 27556.77*** & -17.113*** & 0.2214 \\
& $v_t$ & 0.0002 & 1.4818 & -1.1975 & 0.3375 & 0.55 & 72.84*** & -16.745*** & 0.0167 \\
\midrule
\multirow{3}{*}{DJI} & RV & 0.5665 & 40.1264 & 0.0155 & 1.6341 & 266.12 & 10733939.86*** & -10.191*** & 0.2405 \\
& $r_t$ & 0.0380 & 10.7639 & -13.8418 & 1.0431 & 21.06 & 66928.69*** & -12.722*** & 0.0173 \\
& $v_t$ & 0.0001 & 1.4191 & -1.4339 & 0.3281 & 0.60 & 67.66*** & -15.807*** & 0.0159 \\
\bottomrule
\end{tabular}
\begin{tablenotes}
\footnotesize
    \item[] Note: *, **, and *** denote rejection of the null hypothesis at the 10\%, 5\%, and 1\% significance levels, respectively.

\end{tablenotes}
\end{threeparttable}
}
\end{table*}

Using the OSW-MF-ADCCA algorithm, we compute the time-varying asymmetric Hurst exponents for each index by applying a rolling window of $T=252$ days (i.e., number of trading days in a year) to their respective ($r_t, v_t$) pairs. This process generates the core feature set for our models.

\subsection{Experimental Setup}
\label{subsec:experimental_setup}
Our experimental design is structured to rigorously evaluate the performance and generalizability of the FCOC framework. The primary objective is to forecast the one-day-ahead realized volatility ($RV_{t+1}$) for both the S\&P 500 and DJI indices. To ensure a robust out-of-sample evaluation, each dataset is chronologically divided into training, validation, and test sets according to an approximate 7:1:2 ratio, respectively. This results in three distinct and non-overlapping periods dedicated to model learning, hyperparameter optimization, and final performance assessment.

Prior to training, all input features undergo Min-Max normalization to scale them into a consistent [0, 1] range, a standard procedure to stabilize the learning process. The transformation is defined by the formula:
\begin{equation}
    x' = \frac{x - x_{\min}}{x_{\max} - x_{\min}}
\end{equation}
where $x_{\min}$ and $x_{\max}$ represent the minimum and maximum values of a feature over the training set, respectively. Crucially, for each dataset, these scaling parameters are computed only on its respective training set and are then applied to its validation and testing sets to prevent any data leakage.

For the training process, all models are optimized by minimizing the Mean Squared Error (MSE) loss function using the Adam optimizer. The MSE is chosen as the training objective due to its stability and its property of heavily penalizing large prediction errors. For hyperparameter tuning on the validation set, we adopt a holistic evaluation approach. While all four performance metrics - MSE, Mean Absolute Error (MAE), the Coefficient of Determination (R²), and the Quasi-Likelihood (QLIKE) - are considered to ensure a well-rounded performance, we place a particular emphasis on the QLIKE loss \cite{patton2011volatility}. As a robust loss function for comparing volatility models under noisy conditions, QLIKE serves as a key criterion for final model selection, ensuring that our chosen configurations are both accurate and superior from a risk-management perspective. Key hyperparameters, such as the learning rate (with a primary value of 0.001) and number of hidden neurons, are tuned via the grid search, with a fixed look-back window of 60 days and a batch size of 64.

To ensure the robustness of our findings, we apply the Model Confidence Set (MCS) procedure \cite{hansen2011model} to identify an optimal set of models with predictive advantages, considering that multiple models can be considered the best with the equal predictive accuracy. The MCS test is applied to a pool of models comprising the benchmark models and their variants enhanced with either the standard MF-ADCCA or our proposed OSW-MF-ADCCA features. This comparison allows us to statistically validate the effectiveness of our feature engineering approach before integrating it into the final FCOC framework. At a 25\% significance level, we conduct the procedure with 10,000 bootstrap resamples and a block length of 30 to generate the MCS p-values. 

\section{Empirical Analysis}
\label{sec:empirical_analysis}
To validate the robustness and generalizability of the FCOC framework, we conduct a comprehensive study on forecasting the RV of two of the world's most prominent stock market indices: the S\&P 500 and the DJI. While both represent the mature U.S. market, they differ significantly in their composition and weighting methodology, providing a robust testbed for our methodology. All experiments are implemented in Python 3.10 using the PyTorch 2.0 framework with CUDA 11.8 acceleration on a system equipped with an NVIDIA RTX 3060 GPU, an Intel i7-11800H CPU, and 32GB of RAM.

\subsection{Performance of the Fractal Feature Corrector (FFC)}
\label{subsec:ffc_performance}
We first evaluate the effectiveness of the FFC component by comparing benchmark models against variants enhanced with fractal features. Table \ref{tab:ffc_comparative} details the performance of models augmented with standard MF-ADCCA and our proposed OSW-MF-ADCCA. The results consistently show that enriching models with fractal features improves forecasting accuracy. Notably, the OSW-MF-ADCCA variant, which mitigates spurious fluctuations through overlapping segmentation, generally yields the best performance across all model families and datasets. For example, it boosts the Transformer's R² on the S\&P 500 from 0.1234 to 0.3829, a 210\% increase.

To statistically validate this choice, we conducted a Model Confidence Set (MCS) test, with results presented in Table \ref{tab:mcs_comparative_wide_restructured}. The OSW-MF-ADCCA configuration achieves a p-value of 1.0000 across all scenarios, indicating that it is the sole member of the superior set of models at a 25\% significance level. This confirms the robustness of our feature extraction method and justifies its use as the core of the FFC.

\begin{table*}[!ht]
\centering
\caption{Comparative performance of FFC-enhanced models on the S\&P 500 and DJI indices. Best results within each model family and index are in \textbf{bold}.}
\label{tab:ffc_comparative}
\resizebox{\textwidth}{!}{%
\begin{tabular}{llcccccccc}
\toprule
\multirow{2}{*}{\textbf{Model Family}} & \multirow{2}{*}{\textbf{Method}} & \multicolumn{4}{c}{\textbf{S\&P 500}} & \multicolumn{4}{c}{\textbf{Dow Jones Industrial Average}} \\
\cmidrule(lr){3-6} \cmidrule(lr){7-10}
& & \textbf{R² ($\uparrow$)} & \textbf{QLIKE ($\downarrow$)} & \textbf{MAE ($\downarrow$)} & \textbf{MSE ($\downarrow$)} & \textbf{R² ($\uparrow$)} & \textbf{QLIKE ($\downarrow$)} & \textbf{MAE ($\downarrow$)} & \textbf{MSE ($\downarrow$)} \\
\midrule
\multirow{3}{*}{LSTM} & Benchmark & 0.4460 & 0.2238 & 0.2782 & 0.3030 & 0.4618 & 0.1687 & 0.1936 & 0.1453 \\
                      & MF-ADCCA & 0.4556 & 0.2064 & 0.2784 & 0.2978 & 0.4794 & 0.1602 & 0.1922 & 0.1405 \\
                      & \textbf{OSW-MF-ADCCA} & \textbf{0.4643} & \textbf{0.2066} & \textbf{0.2774} & \textbf{0.2930} & \textbf{0.4968} & \textbf{0.1508} & \textbf{0.1906} & \textbf{0.1358} \\
\midrule
\multirow{3}{*}{GRU} & Benchmark & 0.4335 & 0.2371 & 0.2787 & 0.3099 & 0.3549 & 0.2062 & 0.2113 & 0.1741 \\
                      & MF-ADCCA & 0.4121 & 0.2590 & 0.2850 & 0.3216 & 0.3948 & 0.2095 & 0.2059 & 0.1633 \\
                      & \textbf{OSW-MF-ADCCA} & \textbf{0.4687} & \textbf{0.2056} & \textbf{0.2757} & \textbf{0.2906} & \textbf{0.4051} & \textbf{0.1867} & \textbf{0.2037} & \textbf{0.1606} \\
\midrule
\multirow{3}{*}{Mamba} & Benchmark & 0.4099 & 0.2686 & 0.2861 & 0.3228 & 0.2632 & 0.1602 & 0.2648 & 0.1989 \\
                       & MF-ADCCA & 0.4176 & 0.2616 & 0.2864 & 0.3186 & 0.3197 & 0.1529 & 0.2193 & 0.1836 \\
                       & \textbf{OSW-MF-ADCCA} & \textbf{0.4593} & \textbf{0.2159} & \textbf{0.2811} & \textbf{0.2958} & \textbf{0.4365} & \textbf{0.1498} & \textbf{0.2080} & \textbf{0.1521} \\
\midrule
\multirow{3}{*}{Transformer} & Benchmark & 0.1234 & 0.1993 & 0.3793 & 0.4816 & 0.3700 & 0.1693 & 0.2698 & 0.1701 \\
                             & MF-ADCCA & 0.1411 & 0.2034 & 0.3824 & 0.4719 & -1.0111 & 0.2194 & 0.4321 & 0.5428 \\
                             & \textbf{OSW-MF-ADCCA} & \textbf{0.3829} & \textbf{0.1995} & \textbf{0.3083} & \textbf{0.3375} & \textbf{0.4045} & \textbf{0.1445} & \textbf{0.2188} & \textbf{0.1607} \\
\bottomrule
\end{tabular}%
}
\end{table*}

\begin{table*}[!htbp]
\centering
\caption{Model Confidence Set (MCS) p-values at 25\% Significance Level.}
\label{tab:mcs_comparative_wide_restructured}

\setlength{\tabcolsep}{2.5pt}
\renewcommand{\arraystretch}{0.90}

\resizebox{0.82\textwidth}{!}{%
\begin{tabular}{llccc}
\toprule
\textbf{Index} & \textbf{Model} & \textbf{Benchmark} & \textbf{MF-ADCCA} & \textbf{OSW-MF-ADCCA} \\
\midrule
\multirow{4}{*}{\textbf{S\&P 500}} & LSTM        & 0.0130 & 0.0430 & \textbf{1.0000} \\
                                   & GRU         & 0.0130 & 0.0130 & \textbf{1.0000} \\
                                   & Mamba       & 0.0090 & 0.0060 & \textbf{1.0000} \\
                                   & Transformer & 0.0120 & 0.0130 & \textbf{1.0000} \\
\midrule
\multirow{4}{*}{\textbf{DJI}}      & LSTM        & 0.0250 & 0.0250 & \textbf{1.0000} \\
                                   & GRU         & 0.0480 & 0.0600 & \textbf{1.0000} \\
                                   & Mamba       & 0.0130 & 0.1510 & \textbf{1.0000} \\
                                   & Transformer & 0.2480 & 0.0200 & \textbf{1.0000} \\
\bottomrule
\end{tabular}%
}
\end{table*}

\subsection{Ablation Study and Synergistic Gains of the FCOC Framework}
\label{subsec:ablation_and_synergy}
Having validated the effectiveness of the FFC, we now conduct a comprehensive ablation study to dissect the individual contribution of the Chaotic Oscillation Component (COC) and demonstrate the synergistic power of the full FCOC framework. We compare four configurations: (1) \textbf{Benchmark} (Base Features + ReLU), (2) \textbf{COC-only} (Base Features + COC), (3) \textbf{FFC-only} (Fractal Features + ReLU), and (4) the \textbf{Full FCOC} framework (Fractal Features + COC). The complete results are presented in Table \ref{tab:ablation_study_full}.

The results provide direct and powerful evidence for our central thesis. The "COC-only" configuration demonstrates significant performance gains over the benchmark across all tested architectures, strongly validating our hypothesis of a fundamental complexity mismatch between static processors and dynamic financial signals. This effect is most pronounced in the case of the Transformer on the S\&P 500, where merely replacing the static ReLU with our dynamic COC causes its R² to surge from a dismal 0.1234 to a competitive 0.4413—a dramatic improvement of over 250\%. This proves the standalone novelty and value of the COC.

\begin{figure*}[!ht]
    \centering
    \begin{subfigure}[b]{0.45\textwidth}
        \centering
        \includegraphics[width=\textwidth]{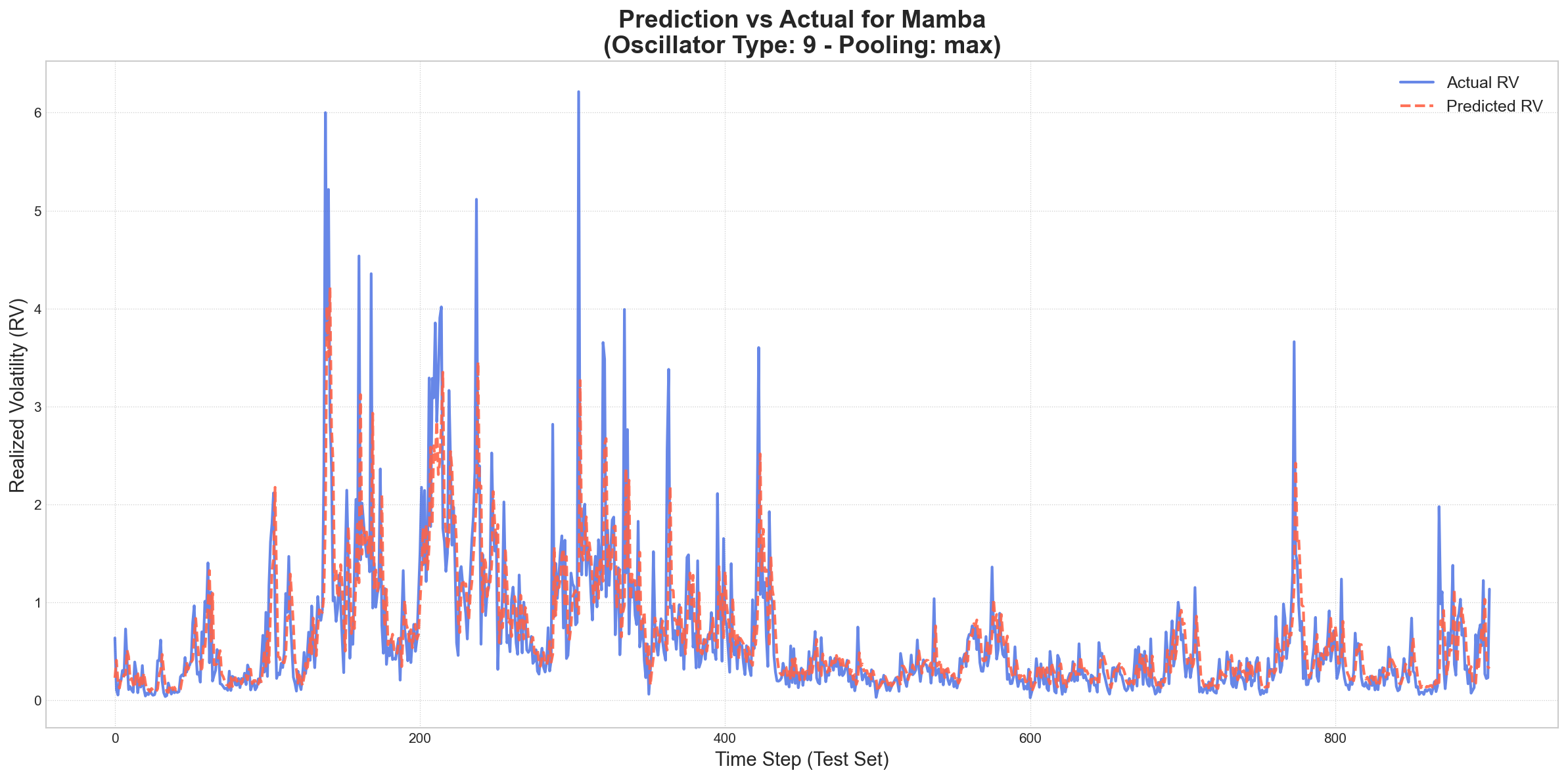}
        \caption{Mamba-FCOC Forecast (S\&P 500)}
        \label{fig:fcoc_mamba_sp500}
    \end{subfigure}
    \hfill
    \begin{subfigure}[b]{0.45\textwidth}
        \centering
        \includegraphics[width=\textwidth]{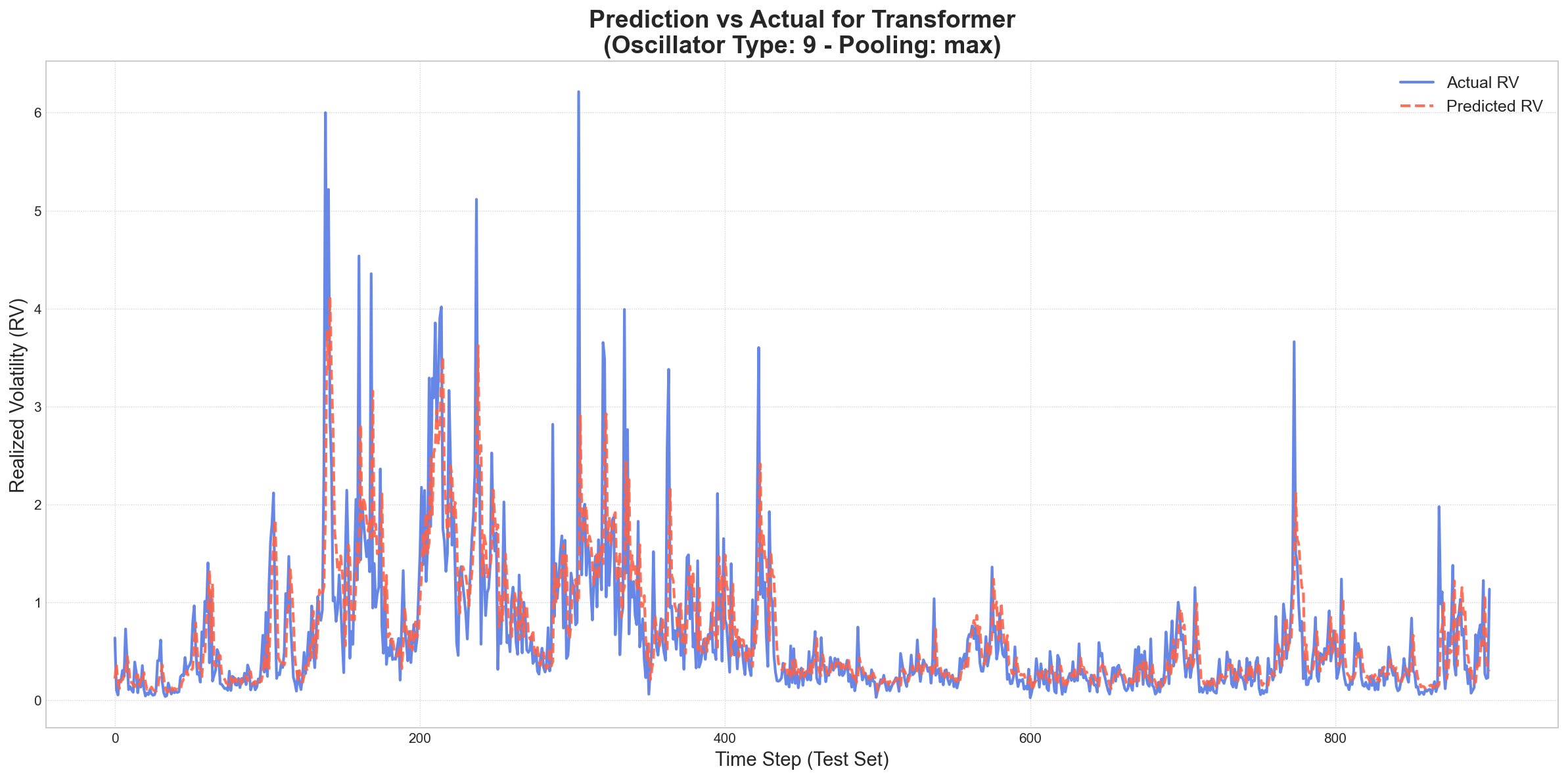} 
        \caption{Transformer-FCOC Forecast (S\&P 500)}
        \label{fig:fcoc_transformer_sp500}
    \end{subfigure}
    \vskip\baselineskip
    \begin{subfigure}[b]{0.45\textwidth}
        \centering
        \includegraphics[width=\textwidth]{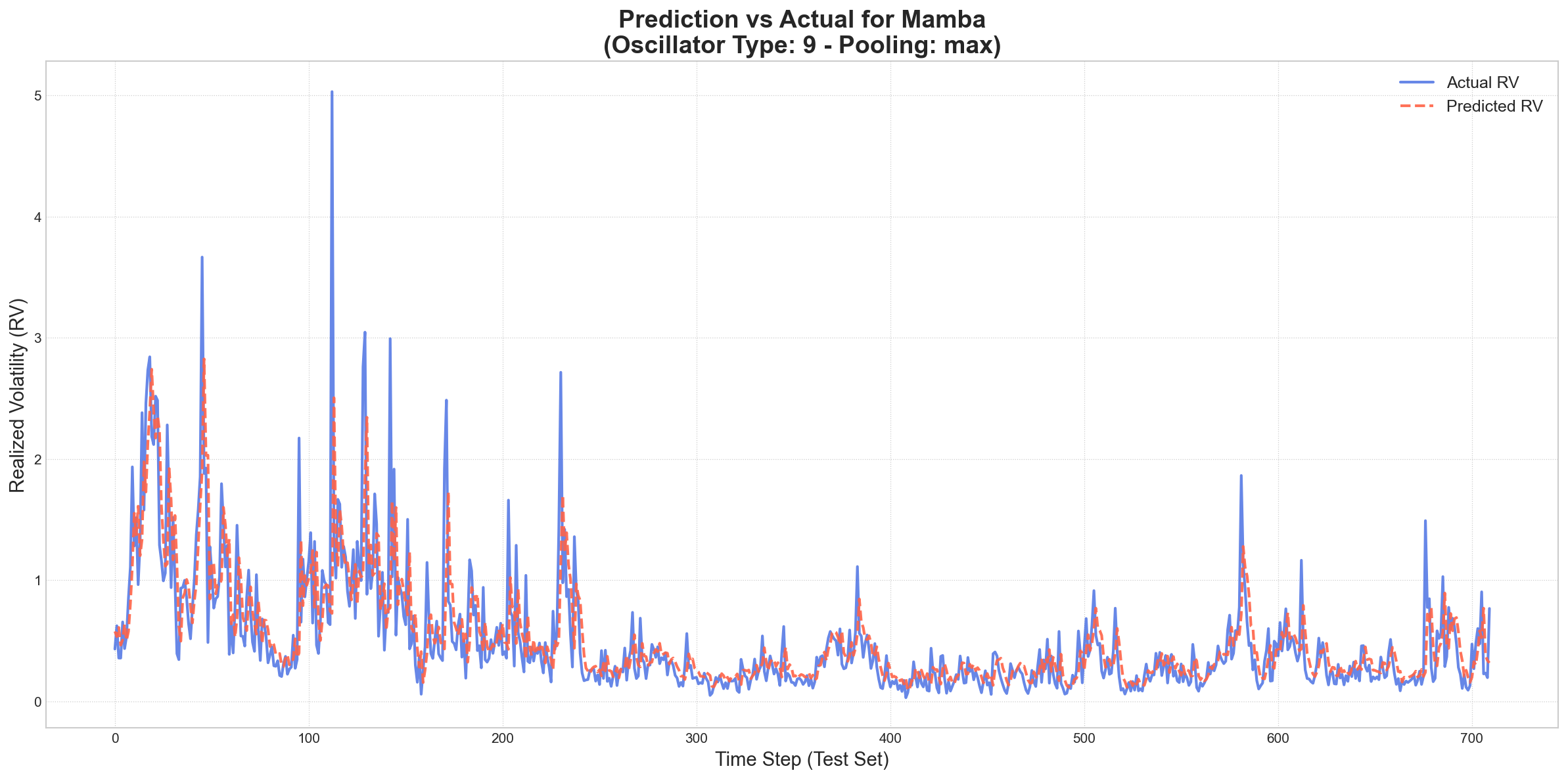}
        \caption{Mamba-FCOC Forecast (DJI)}
        \label{fig:fcoc_mamba_dji}
    \end{subfigure}
    \hfill
    \begin{subfigure}[b]{0.45\textwidth}
        \centering
        \includegraphics[width=\textwidth]{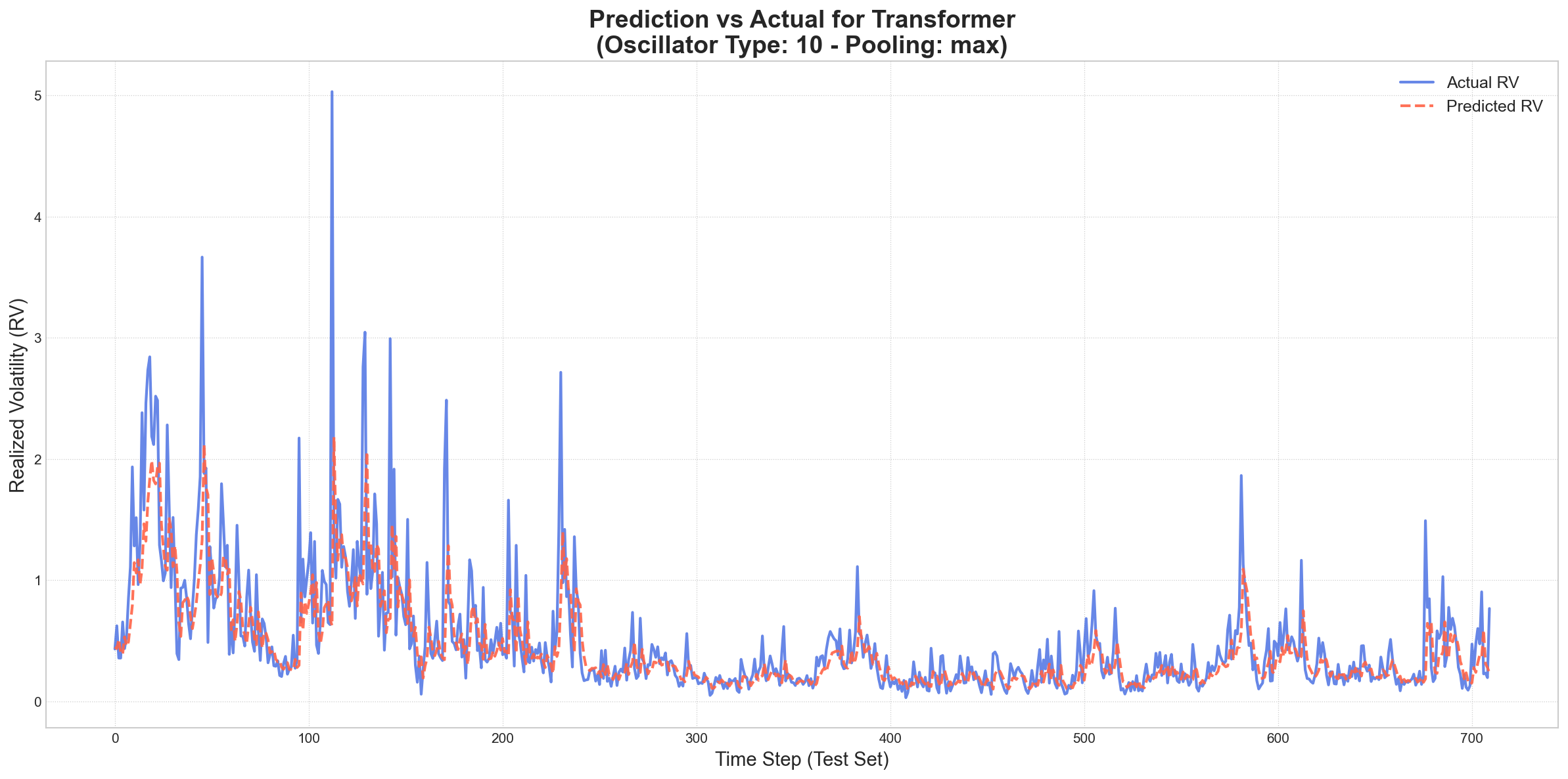} 
        \caption{Transformer-FCOC Forecast (DJI)}
        \label{fig:fcoc_transformer_dji}
    \end{subfigure}
    \caption{Comparative out-of-sample forecasting performance for representative FCOC models on the S\&P 500 and DJI test sets.}
    \label{fig:fcoc_forecasts_comparative}
\end{figure*}

Ultimately, the full FCOC framework consistently delivers the best overall performance, revealing a powerful synergistic effect. In nearly every case, the full framework (4) outperforms not only the benchmark (1) but also both single-component configurations (2 and 3). For example, on the DJI, the FCOC-Mamba achieves an R² of 0.5066, substantially higher than both the FFC-only (0.4365) and COC-only (0.4913) versions. This confirms that the highest-fidelity signals from the FFC are most effectively leveraged by the commensurately complex processor of the COC, validating our co-driven design philosophy. The visual forecasts in Figure \ref{fig:fcoc_forecasts_comparative} further corroborate the superior accuracy of the full FCOC models.

\begin{table*}[!htbp]
\centering
\caption{Comprehensive Ablation Study of FCOC Framework Components on S\&P 500 and DJI Test Sets.}
\label{tab:ablation_study_full}
\resizebox{\textwidth}{!}{%
\begin{threeparttable}
\begin{tabular}{llcccc|cccc}
\toprule
\multirow{2}{*}{\textbf{Dataset}} & \multirow{2}{*}{\textbf{Model}} & \multicolumn{4}{c}{\textbf{(1) Benchmark}} & \multicolumn{4}{c}{\textbf{(2) COC-only}} \\
\cmidrule(lr){3-6} \cmidrule(lr){7-10}
& & \textbf{R²} ($\uparrow$) & \textbf{QLIKE} ($\downarrow$) & \textbf{MAE} ($\downarrow$) & \textbf{MSE} ($\downarrow$) & \textbf{R²} ($\uparrow$) & \textbf{QLIKE} ($\downarrow$) & \textbf{MAE} ($\downarrow$) & \textbf{MSE} ($\downarrow$) \\
\midrule
\multirow{4}{*}{\textbf{S\&P 500}} 
& LSTM & 0.4460 & 0.2238 & 0.2782 & 0.3030 & 0.4738 & 0.1920 & 0.2713 & 0.2591 \\
& GRU & 0.4335 & 0.2371 & 0.2787 & 0.3099 & 0.4633 & 0.2032 & 0.2652 & 0.2643 \\
& Mamba & 0.4099 & 0.2686 & 0.2861 & 0.3228 & 0.4726 & 0.2008 & 0.2668 & 0.2597 \\
& Transformer & 0.1234 & 0.1993 & 0.3793 & 0.4816 & 0.4413 & 0.1912 & 0.2866 & 0.2751 \\
\midrule
\multirow{4}{*}{\textbf{DJI}} 
& LSTM & 0.4618 & 0.1687 & 0.1936 & 0.1453 & 0.4849 & 0.1512 & 0.1941 & 0.1401 \\
& GRU & 0.3549 & 0.2062 & 0.2113 & 0.1741 & 0.4339 & 0.1953 & 0.2008 & 0.1540 \\
& Mamba & 0.2632 & 0.1602 & 0.2648 & 0.1989 & 0.4913 & 0.1506 & 0.1959 & 0.1384 \\
& Transformer & 0.3700 & 0.1693 & 0.2698 & 0.1701 & 0.4717 & 0.1595 & 0.1990 & 0.1437 \\
\midrule
\midrule
\multirow{2}{*}{\textbf{Dataset}} & \multirow{2}{*}{\textbf{Model}} & \multicolumn{4}{c}{\textbf{(3) FFC-only}} & \multicolumn{4}{c}{\textbf{(4) Full FCOC Framework}} \\
\cmidrule(lr){3-6} \cmidrule(lr){7-10}
& & \textbf{R²} ($\uparrow$) & \textbf{QLIKE} ($\downarrow$) & \textbf{MAE} ($\downarrow$) & \textbf{MSE} ($\downarrow$) & \textbf{R²} ($\uparrow$) & \textbf{QLIKE} ($\downarrow$) & \textbf{MAE} ($\downarrow$) & \textbf{MSE} ($\downarrow$) \\
\midrule
\multirow{4}{*}{\textbf{S\&P 500}} 
& LSTM & 0.4643 & 0.2066 & 0.2774 & 0.2930 & \textbf{0.4950} & \textbf{0.1872} & 0.2822 & \textbf{0.2774} \\
& GRU & 0.4687 & 0.2056 & 0.2757 & 0.2906 & \textbf{0.4848} & \textbf{0.1906} & 0.2877 & \textbf{0.2831} \\
& Mamba & 0.4593 & 0.2159 & 0.2811 & 0.2958 & \textbf{0.4762} & \textbf{0.1899} & 0.2921 & \textbf{0.2865} \\
& Transformer & 0.3829 & 0.1995 & 0.3083 & 0.3375 & \textbf{0.4733} & \textbf{0.1920} & 0.2980 & \textbf{0.2881} \\
\midrule
\multirow{4}{*}{\textbf{DJI}} 
& LSTM & 0.4968 & 0.1508 & 0.1906 & 0.1358 & \textbf{0.5061} & \textbf{0.1400} & 0.1946 & \textbf{0.1333} \\
& GRU & 0.4051 & 0.1867 & 0.2037 & 0.1606 & \textbf{0.4621} & \textbf{0.1545} & 0.2053 & \textbf{0.1452} \\
& Mamba & 0.4365 & 0.1498 & 0.2080 & 0.1521 & \textbf{0.5066} & \textbf{0.1384} & 0.2031 & \textbf{0.1332} \\
& Transformer & 0.4045 & 0.1445 & 0.2188 & 0.1607 & \textbf{0.4851} & \textbf{0.1590} & \textbf{0.1919} & \textbf{0.1390} \\
\bottomrule
\end{tabular}
\begin{tablenotes}
\footnotesize
    \item[*] Note: Best results for each metric within each (Dataset, Model) group are highlighted in \textbf{bold}.
\end{tablenotes}
\end{threeparttable}
}
\end{table*}

\subsection{Discussion}
\label{subsec:discussion}
The comprehensive empirical results do more than simply demonstrate superior forecasting accuracy; they provide deep insights into the architectural synergies of our co-driven framework and reveal fundamental limitations in current deep learning approaches to financial forecasting. The framework's impact is perhaps most vividly demonstrated by resolving what we term the Transformer Paradox. While the benchmark Transformer fails on noisy financial data due to its unconstrained attention mechanism, the FCOC framework orchestrates a dramatic recovery. The ablation study proves unequivocally that the Chaotic Oscillation Component (COC) is the decisive factor, as its introduction alone is sufficient to fix the Transformer's core weakness by providing a dynamic processing unit whose complexity matches the signal. While the framework serves as a corrective measure for such architectures, for models with strong sequential priors like LSTM and Mamba, it acts as a powerful empowerment layer. Here, the synergy of our framework's two pillars shines: the Fractal Feature Corrector (FFC) provides a direct, high-fidelity signal of the market's memory state, which is then processed by the COC's dynamic engine. The spectacular 92.5\% R² improvement for the FCOC-Mamba model on the DJI dataset is the definitive evidence of this co-driven synergy, unlocking the full potential of advanced architectures.

Beyond these performance gains, our findings point toward deeper principles and promising new research frontiers. An intriguing finding is the models' architectural affinity for specific chaotic dynamics, such as the preference of S\&P 500 models for the multi-modal T9 oscillator. We speculate that this is not arbitrary, but rather evidence of a "dynamic resonance" between the model's internal structure and the data's topological properties. This suggests a future research direction beyond one-size-fits-all activations toward architecture-dynamics co-design. From a practical standpoint, the nuanced behavior of the error metrics further underscores the framework's value. The marked improvements in MSE and QLIKE, even with slight degradation in MAE in some cases, are not a trade-off but a desirable feature for risk management. It demonstrates that our framework excels at mitigating catastrophic risk during market turmoil by heavily penalizing large errors, a highly favorable characteristic for any real-world financial application.

\section{Conclusion}
\label{sec:conclusion}
This paper introduces the FCOC framework, a novel architecture designed to address the dual bottlenecks of feature fidelity and model responsiveness in volatility forecasting. By synergistically combining our novel Fractal Feature Corrector with a dynamic Chaotic Oscillation Component, our framework significantly enhances the predictive power of advanced deep learning models. Our comprehensive empirical study on the S\&P 500 and DJI demonstrates the framework's robustness and generalizability. The findings reveal that FCOC not only empowers sequential models like LSTM by providing explicit market-state information but also offers a path to resolving the Transformer Paradox by equipping its attention mechanism with a crucial inductive bias. This work validates a new co-driven paradigm, showing that the synergy between superior theoretical features and dynamic internal processors is a key factor in unlocking the full potential of deep learning in complex financial environments. Future research should focus on both deepening the theoretical underpinnings of this approach---for instance, by using nonlinear time series and topological data analysis to quantitatively compare the market's attractor reconstruction with the oscillators' phase space structures---and extending its practical applications to extreme settings like cryptocurrency markets and domains such as algorithmic trading and portfolio optimization.

\section*{Declaration of Competing Interest}
The authors declare that they have no known competing financial interests or personal relationships that could have appeared to influence the work reported in this paper.

 \section*{CRediT authorship contribution statement}
 \textbf{Yilong Zeng:} Conceptualization, Methodology, Software, Validation, Formal analysis, Investigation, Writing – Original Draft. 
 \textbf{Boyan Tang:} Methodology, Software, Data Curation.
 \textbf{Xuanhao Ren:} Visualization, Investigation.
 \textbf{Sherry Zhefang Zhou:} Conceptualization, Supervision, Writing – Review \& Editing, Funding acquisition.
 \textbf{Jianghua Wu:} Supervision, Project administration.
 \textbf{Raymond Lee:} Conceptualization, Supervision, Writing – Review \& Editing, Funding acquisition.

 \section*{Acknowledgments}
  This work was supported in part by the Guangdong Provincial Key Laboratory of IRADS (2022B1212010006) and the Shenzhen Research Institute of Big Data (J00220240006).

\section*{Data availability statement}
The data that support the findings of this study are available from Wind. Restrictions apply to the availability of these data, which are used under license for this study. Data are available from https://www.wind.com.cn/ 

\bibliographystyle{elsarticle-num} 
\bibliography{my_references} 

\end{document}